\documentclass{ws-procs9x6}

\begin{document}

\title{Shape-invariance and Exactly Solvable Problems in Quantum
  Mechanics}   

\author{A.~B. BALANTEKIN\\}

\address{University of Wisconsin, Department of Physics \\
Madison, WI  53706,  USA \\ 
E-mail: baha@nucth.physics.wisc.edu}

\maketitle

\abstracts{Algebraic approach to the integrability condition
  called shape invariance is briefly reviewed. Various applications of
  shape-invariance available in the literature are listed. A class of
  shape-invariant bound-state problems which represent two-level
  systems are examined. These generalize the Jaynes-Cummings
  Hamiltonian. Coherent states associated with shape-invariant systems
  are discussed. For the case of quantum harmonic oscillator the
  decomposition of identity for these coherent states is given. This
  decomposition of identity utilizes Ramanujan's integral extension of
  the beta function.}

\section{Introduction}\label{1A}

The technique of factorization is a widely-used method to find
eigenvalues and eigenvectors of quantum mechanical Hamiltonians. The
factorization method was most recently utilized in the context of
supersymmetric quantum mechanics \cite{Witten:nf,Cooper:1994eh}. In
this method the Hamiltonian, after subtracting the ground state
energy, 
is written as the product of an operator $\hat{A}$ and its Hermitian
conjugate, $\hat{A}^{\dagger}$:
\begin{equation}
  \label{1}
  \hat{H} - E_0 = \hat{A}^{\dagger}\hat{A}, 
\end{equation}
where $E_0$ is the ground state energy. With this definition the
ground state wavefunction in supersymmetric quantum mechanics
is annihilated by the operator $\hat{A}$:  
\begin{equation}
  \label{2}
  \hat{A}| \psi_0 \rangle = 0.
\end{equation}
The Hamiltonian in Eq. (\ref{1}) is called shape-invariant
\cite{Gendenshtein:vs} if the condition
\begin{equation}
  \label{3}
  \hat{A}(a_1)\hat{A}^{\dagger}(a_1) = \hat{A}^{\dagger}(a_2)
  \hat{A}(a_2) + R(a_1) 
\end{equation}
is satisfied. In Eq. (\ref{3}) $a_1, a_2, \cdots$ represent the
parameters of the Hamiltonian. (The original Hamiltonian has the
parameter $a_1$, the transformed Hamiltonian has $a_2$ and so on. 
The parameter $a_2$ is a
function of the parameter $a_1$ and the
remainder $R(a_1)$ is
independent of the dynamical variables of the problem.

Shape-invariance problem was formulated in algebraic terms in
Ref. [4]. To introduce this formalism we define an operator
which transforms the parameters of the potential:
\begin{equation}
  \label{4}
\hat{T}(a_1) { O}(a_1) \hat{T}^{-1}(a_1) = { O}(a_2).
\end{equation}
Introducing new operators 
\begin{eqnarray}
  \label{5}
  \hat{B}_+ &=& \hat{A}^{\dagger}(a_1) \hat{T} (a_1) \nonumber \\
 \hat{B}_- = \hat{B}_+^{\dagger} &=& \hat{T}^{\dagger}(a_1) \hat{A}
 (a_1). 
\end{eqnarray}
one can show that the Hamiltonian can be written as
\begin{equation}
  \label{6}
\hat{H} - E_0 = \hat{A}^{\dagger}\hat{A} = \hat{B}_+ \hat{B}_- .
\end{equation}
Using the definition given in Eq. (\ref{5}), the shape-invariance
condition of Eq. (\ref{3}) takes the form
\begin{equation}
  \label{7}
[ \hat{B}_- , \hat{B}_+ ] = R(a_0),
\end{equation}
where $R(a_0)$ is defined via
\begin{equation}
  \label{8}
R(a_n) = \hat{T}(a_1) R(a_{n-1})\hat{T}^{\dagger}(a_1).
\end{equation}
One can show that 
\begin{equation}
  \label{9}
\hat{B}_- | \psi_0 \rangle = 0,
\end{equation}
\begin{equation}
  \label{10}
[\hat{H}, \hat{B}_+^n ] = (R(a_1)+R(a_2)+ \cdot \cdot + R(a_n))
  \hat{B}_+^n ,
\end{equation}
and 
\begin{equation}
  \label{11}
[\hat{H}, \hat{B}_-^n ] = - \hat{B}_-^n (R(a_1)+R(a_2)+ \cdot \cdot
  + R(a_n))\,, 
\end{equation}
i.e. $\hat{B}_+^n |\psi_0 \rangle$ is an eigenstate of the
Hamiltonian
with
the eigenvalue $R(a_1)+R(a_2)+ \cdot \cdot + R(a_n)$.
The normalized wavefunction is
\begin{equation}
  \label{12}
|\psi_n \rangle = \frac{1}{\sqrt{R(a_1)+ \cdot \cdot +
R(a_n)}} \hat{B}_+ \cdot \cdot \frac{1}{\sqrt{R(a_1)+
R(a_2)}} \hat{B}_+  \frac{1}{\sqrt{R(a_1)}}\hat{B}_+
| \psi_0 \rangle.
\end{equation}

The algebra is given by the commutators 
\begin{equation}
  \label{13}
[ \hat{B}_- , \hat{B}_+ ] = R(a_0),
\end{equation}
\begin{equation}
  \label{14}
[ \hat{B}_+ , R(a_0) ] = (R(a_1) - R(a_0)) \hat{B}_+,
\end{equation}
and 
\begin{equation}
  \label{15}
[ \hat{B}_+ , (R(a_1)-R(a_0))\hat{B}_+ ] = \{ (R(a_2) -
  R(a_1))-(R(a_1) - R(a_0))\} \hat{B}_+^2,
\end{equation}
and so on. In general there are an infinite number of such commutation
relations. If the quantities $R(a_n)$ satisfy certain relations one of
the commutators in this series may vanish. For such a situation the
commutation relations obtained up to that point plus their complex
conjugates form a Lie algebra with a finite number of elements.

In the shape-invariant problem the parameters of the Hamiltonian are
viewed as auxiliary dynamical variables. One can imagine an
alternative approach of classifying some of the dynamical variables as
``parameters''. An example of this is provided by the supersymmetric
approach to the spherical Nilsson model of single particle states
\cite{Balantekin:1992qp}. The Nilsson Hamiltonian is given by
\begin{equation}
  \label{16}
H = \sum_i a^\dagger_i a_i - 2 k {\bf L . S} + k \nu {\bf L}^2 .
\end{equation}
The superalgebra $Osp(1/2)$ is the dynamical symmetry algebra 
of this problem \cite{Balantekin:1984hf}. Introducing the odd
generator of this superalgebra 
\begin{equation}
  \label{17}
F^\dagger = \sum_i \sigma_i  a^\dagger_i
\end{equation}
one can show that the ``Hamiltonians''
\begin{equation}
  \label{18}
H_1 = F^\dagger F = \sum_i a^\dagger_i a_i - {\bf \sigma . L}
\end{equation}
and
\begin{equation}
  \label{19}
H_2 = F F^\dagger =  \sum_i a_i a^\dagger_i + {\bf \sigma . L}
\end{equation}
can be considered as supersymmetric partners of each other
\cite{Balantekin:1984hf}. The
shape-invariance condition of Eq. (\ref{3}) can be written as
\begin{equation}
  \label{20}
F F^\dagger = F^\dagger F + R,
\end{equation}
where the remainder is
\begin{equation}
  \label{21}
R= {\bf \sigma . L} - 3/4 ,
\end{equation}
i.e. in this example the radial variables are considered as the main
dynamical variables and the angular variables are considered as the
``auxiliary parameters''.

A number of applications of shape-invariance are available in the
literature. These include i) Quantum tunneling through supersymmetric
shape-invariant potentials \cite{Aleixo:1999cx}; ii) Study of neutrino
propagation through shape-invariant electron densities
\cite{Balantekin:1997jp}; iii) Exploration of the relationship between
algebraic techniques of Gaudin developed to deal with many-spin
systems, quasi-exactly solvable potentials,  
and shape-invariance \cite{gernot};              
iv) Investigation of coherent states for
shape-invariant potentials \cite{Balantekin:1998wj,Aleixo:2002sa}; and
v) As attempts to devise exactly solvable coupled-channel problems,
generalization of Jaynes-Cummings type Hamiltonians to shape-invariant
systems \cite{Aleixo:2000ub,Aleixo:2001jm}. In this article we focus
on the last two applications. 

\section{A Generalized Jaynes-Cummings Hamiltonian For Shape-Invariant
  Systems}

Attempts were made to generalize
supersymmetric quantum mechanics and the concept of shape-invariance
to coupled-channel problems \cite{amado,das}. 
In general it is not easy to find exact solutions to
coupled-channels problems. In the coupled-channels case a general
shape-invariance is only possible in the limit where the
superpotential is separable \cite{das} which corresponds to the
well-known sudden approximation in the coupled-channels problem
\cite{Balantekin:1997yh}. However it is possible to solve 
a class of shape-invariant coupled-channels problems which correspond
to the 
generalization of the Jaynes-Cummings Hamiltonian \cite{jc} widely
used in atomic physics to describe a two-level atom interacting with
photons:  
\begin{equation}
  \label{22}
\hat H_{\rm JC} = \omega_0 a^\dagger a + \omega \sigma_3 + \Omega
\left( 
\sigma_+\hat a + \sigma_-\hat a^\dagger\   \right) .
\end{equation}

The shape-invariant generalization of the Jaynes-Cummings Hamiltonian
is \cite{Aleixo:2000ub}: 
\begin{equation}
  \label{23}
\hat H_{\rm SUSYJC} = \hat A^\dagger\hat A + {1\over 2}\left[\hat
  A,\hat A^\dagger\right]\left(\sigma_3+1\right) +
\sqrt{\hbar\Omega}\left(\sigma_+\hat A+\sigma_-\hat A^\dagger\right)\,.
\end{equation}
To find the eigenvalues of the Hamiltonian in Eq. (\ref{23}) we
introduce the operator 
\begin{equation}
  \label{24}
\hat S = \sigma_+\hat A + \sigma_-\hat A^\dagger\,
\end{equation}
the square of which can be written as 
\begin{equation}
  \label{25}
\hat S^2 = \left[ \matrix{\hat T & 0\cr 0 & \pm 1 \cr}\right] \left[
\matrix{\hat B_-\hat B_+ & 0\cr 0 & \hat B_+\hat B_- \cr}\right]
\left[ \matrix{\hat T^\dagger & 0\cr 0 & \pm 1 \cr}\right]\,.
\end{equation}
We now introduce the states
\begin{equation}
  \label{26}
\mid \Psi_m\rangle_\pm =  \frac{1}{\sqrt{2}}\left[ \matrix{\hat T &
0\cr 0 & \pm 1 \cr}\right] \left[ \matrix{\mid m\rangle\cr \mid
m+1\rangle \cr}\right], \> m=0,1,2, \cdots \label{eqpsm+-}, 
\end{equation}
where $\mid m\rangle$ is the eigenstate of the shape-invariant
Hamiltonian $\hat A^\dagger\hat A$ with eigenvalue
$\varepsilon_{m}$. It can be shown that the states in Eq. (\ref{26}) 
are the eigenstates of the operator $\hat S$:
\begin{equation}
  \label{27}
\hat S \mid \Psi_m\rangle_\pm = \sqrt{\varepsilon_{m+1}} \mid
\Psi_m\rangle_\pm\,.
\end{equation}

Since the Hamiltonian of Eq. (\ref{23}) can be written as
\begin{equation}
  \label{28}
\hat H_{\rm SUSYJC} = \hat S^2 + \sqrt{\hbar\Omega}\,\hat S\,,
\end{equation}
it has the eigenvalue spectrum 
\begin{equation}
  \label{29}
\hat H_{\rm SUSYJC} \mid\Psi_m\rangle_\pm = \left( \varepsilon_{m+1} 
\pm\sqrt{\hbar\Omega} \sqrt{\varepsilon_{m+1}}\right)\mid
\Psi_m\rangle_\pm ,
\end{equation}
for all states except the ground state which is given by 
\begin{equation}
  \label{30}
\mid \Psi_0\rangle  =  \left[ \matrix{ 0 \cr \mid 0 \rangle
\cr}\right], 
\end{equation}
where  $\mid 0 \rangle$ is the ground state of $\hat A^\dagger\hat
A$. The Hamiltonian $H_{\rm SUSYJC}$ has an eigenvalue $0$ on the
state given in Eq. (\ref{30}). A variant of the usual Jaynes-Cummings 
Model takes the coupling between matter and the radiation to depend on
the intensity of the electromagnetic field. This variant can also be
generalized to shape-invariant systems \cite{Aleixo:2001jm}.

\section{Coherent States for the Quantum Oscillator and Ramanujan
  Integrals}

\subsection{Quantum Oscillator as a Shape-invariant Potential}

One class of shape-invariant potentials are
reflectionless potentials with an infinite number of bound states,
also called self-similar potentials \cite{shabat,Spiridonov:md}.
Shape-invariance of such potentials were studied in detail in Refs.
[20] and [21]. For such potentials the parameters are related by a
scaling: 
\begin{equation}
a_n = q^{n-1}a_1\,.
\label{31}
\end{equation}
For the simplest case studied in Ref. [21] the
remainder of Eq.~(\ref{3}) is given by
\begin{equation}
  \label{33}
R(a_1)= ca_1 \,,
\end{equation}
which corresponds to the quantum harmonic oscillator. 
Introducing the operators 
\begin{equation}
  \label{34}
\hat S_+ = \sqrt{q}\hat B_+R(a_1)^{-1/2}
\end{equation}
and
\begin{equation}
  \label{35}
\hat S_- = (\hat S_+)^\dagger = \sqrt{q} R(a_1)^{-1/2}\hat B_-\,,
\end{equation}
one can write the Hamiltonian of the quantum harmonic oscillator as 
\begin{equation}
  \label{36}
\hat H - E_0 = R(a_1) \hat S_+ \hat S_- \, .
\end{equation}
This Hamiltonian has the energy eigenvalues 
\begin{equation}
  \label{37}
E_n = R(a_1) \frac{1-q^n}{1-q} \,,
\end{equation}
and the eigenvectors 
\begin{equation}
  \label{38}
\mid  n\rangle = \sqrt{\frac{(1-q)^n}{(q;q)_n}} (\hat S_+)^n \mid
  0\rangle \, .
\end{equation}
In writing down Eq. (\ref{38}) we used the $q$-shifted factorial
defined as 
\begin{equation}
  \label{39}
(z;q)_0 = 1, \,\, (z;q)_n = \prod_{j=0}^{n-1}(1-zq^j)\,, \,\,\,\, n =
1,2,\dots
\end{equation}

\subsection{Coherent States for Shape-Invariant Systems}

Coherent states for shape-invariant potentials were introduced in
Refs. [9] and [22]. (For a description of an 
alternative approach see Ref. [23] and references therein). 
Following the
definitions in Eqs. (\ref{5}) and (\ref{6}) (with $E_0=0$) we introduce
the operator 
\begin{equation}
  \label{40}
\hat H^{-1}\hat B_+ =  \hat B_-^{-1}, \>\>\>\> (\hat B_-\hat B_-^{-1}
=  1) .
\end{equation}
The coherent state can be defined as \cite{Balantekin:1998wj}: 
\begin{equation}
  \label{41}
\mid z\rangle = \sum_{n=0}^{K} \left( z f[R(a_1)] \hat B_-^{-1}
  \right)^n  \mid 0\rangle ,
\end{equation}
where $f(t)$ is an arbitrary function. This state can explicitly be
written as                        
\begin{eqnarray}
\label{42}
   \mid z\rangle &=&  \mid 0\rangle + z \> \>
   \frac{f[R(a_1)]}{\sqrt{R(a_1)}} \mid 1 \rangle \nonumber \\ &+& z^2
   \frac{f[R(a_1)]f[R(a_2)]}{\sqrt{R(a_2)[R(a_1)+R(a_2)]}} \mid 2
   \rangle \nonumber \\
   &+& z^3
   \frac{f[R(a_1)]f[R(a_2)]f[R(a_3)]}{\sqrt{R(a_3)[R(a_2)+R(a_3)]
[R(a_3)+R(a_2)+R(a_1)]}}
   \mid 3 \rangle \nonumber \\ &+& \cdots  
\end{eqnarray}
where we used the normalized eigenstates of the operator $\hat H$: 
\begin{equation}
  \label{43}
\mid n\rangle = \left[ \hat H^{-1/2}\hat B_+ \right]^n\mid 0\rangle\,.
\end{equation} 
In a similar way to the coherent states for the ordinary harmonic
oscillator the coherent state in Eq. (\ref{41}) is an
eigenstate of the operator $\hat B_-$:
\begin{equation}
\label{44a}
\hat B_-\mid z\rangle = z f[R(a_0)] \mid z\rangle.
\end{equation}

\subsection{q-Coherent States:}

To derive the overcompleteness relation of q-coherent states here we
follow the proof given in Ref. [11]. An  
alternative, but equivalent, derivation was given in Ref. [24]. 
To introduce the coherent states for the q-oscillator we take the
arbitrary function in Eq. (\ref{41}) to be 
\begin{equation}
  \label{44}
f[R(a_n)] =  R(a_n) .
\end{equation}
The resulting coherent states are 
\begin{equation}
  \label{45}
\mid z\rangle = \sum_{n=0}^{\infty} \frac{(1-q)^{n/2}}{\sqrt{(q;q)_n}}
q^{n(n-1)/4}  \sqrt{\left[R(a_1)\right]^n} z^n \mid n \rangle.
\end{equation}
Further introducing the auxiliary variable 
\begin{equation}
  \label{46}
\zeta = \frac{\sqrt{(1-q)}}{\sqrt{q}} \sqrt{R(a_1)} z
\end{equation}
these coherent states take the form
\begin{equation}
  \label{47}
\mid \zeta \rangle = \sum_{n=0}^{\infty}
  \frac{q^{n(n+1)/4}}{\sqrt{(q;q)_n}} \zeta^n  \mid n \rangle.
\end{equation}
The overcompleteness of these coherent states can easily be proven
using the integral  
\begin{equation}
  \label{48}
\int_0^{\infty} dt \frac{t^n}{(-t;q)_{\infty}} =
  \frac{(q;q)_n}{q^{n(n+1)/2}} (-\log q) .
\end{equation} 
This integral was proven by Ramanujan in an attempt to
generalize integral definition of the beta function \cite{raman}. (An
elementary proof is given by Askey in Ref. [26]). Using
Eq. (\ref{48}) the overcompleteness relation of the coherent states 
in Eq. (\ref{47}) can be obtained in a straightforward way: 
\begin{equation}
  \label{49}
 I= \int \frac{d\zeta d\zeta^*}{2\pi i} \frac{1}{(-\log q)}
  \frac{1}{(-\mid \zeta \mid^2;q)_{\infty}}  \mid \zeta \rangle
  \langle \zeta \mid = \hat 1 \,.
\end{equation}
This overcompleteness relation could be useful to write down
coherent-state path integrals for quantum harmonic oscillator.

I would like to express my gratitude to my collaborators G. Akemann,
A. Aleixo, J. Beacom, and M.A. Candido Ribeiro who contributed to
various aspects of the work reported here.  
This work was supported in part by the U.S. National Science
Foundation Grants No. INT-0070889, PHY-0070161, and PHY-0244384.


\begin{thebibliography}{0}


\bibitem{Witten:nf}
E.~Witten,
{\it Nucl.\ Phys.} {\bf B188}, 513 (1981).

\bibitem{Cooper:1994eh}
F.~Cooper, A.~Khare and U.~Sukhatme,
{\it Phys.\ Rept.}  {\bf 251}, 267 (1995)
[arXiv:hep-th/9405029].

\bibitem{Gendenshtein:vs}
L.~E.~Gendenshtein,
%
{\it JETP Lett.}  {\bf 38}, 356 (1983)
[Pisma Zh.\ Eksp.\ Teor.\ Fiz.\  {\bf 38}, 299 (1983)].

\bibitem{Balantekin:1997mg}
A.~B.~Balantekin,
{\it Phys.\ Rev.\ A} {\bf 57}, 4188 (1998)
[arXiv:quant-ph/9712018].

\bibitem{Balantekin:1992qp}
A.~B.~Balantekin, O.~Castanos and M.~Moshinsky,
{\it Phys.\ Lett.\ B} {\bf 284}, 1 (1992).

\bibitem{Balantekin:1984hf}
A.~B.~Balantekin,
Annals Phys.\  {\bf 164}, 277 (1985).

\bibitem{Aleixo:1999cx}
A.~N.~Aleixo, A.~B.~Balantekin and M.~A.~Candido Ribeiro,
{\it J.\ Phys.\ A} {\bf 33}, 1503 (2000)
[arXiv:quant-ph/9910051].

\bibitem{Balantekin:1997jp}
A.~B.~Balantekin,
{\it Phys.\ Rev.\ D} {\bf 58}, 013001 (1998)
[arXiv:hep-ph/9712304]; 
see also 
A.~B.~Balantekin and J.~F.~Beacom,
{\em Phys.\ Rev.} D {\bf 54}, 6323 (1996)
[arXiv:hep-ph/9606353].

\bibitem{gernot}
A.B. Balantekin, to be published in the {\em Proceedings of the
  Ettore Majorana Workshop on Symmetries in Nuclear Structure, March
  2003}, A. Vitturi and R. Casten, Editors (World Scientific, 2003); 
G. Akemann and A.B. Balantekin, in preparation.


\bibitem{Balantekin:1998wj}
A.~B.~Balantekin, M.~A.~Candido Ribeiro and A.~N.~Aleixo,
{\it J.\ Phys.\ A} {\bf 32}, 2785 (1999)
[arXiv:quant-ph/9811061].

\bibitem{Aleixo:2002sa}
A.~N.~Aleixo, A.~B.~Balantekin and M.~A.~Candido Ribeiro,
%
{\it J.\ Phys.\ A} {\bf 35}, 9063 (2002)
[arXiv:math-ph/0209033].

\bibitem{Aleixo:2000ub}
A.~N.~Aleixo, A.~B.~Balantekin and M.~A.~Candido Ribeiro,
{\it J.\ Phys.\ A} {\bf 33}, 3173 (2000)
[arXiv:quant-ph/0001049].

\bibitem{Aleixo:2001jm}
A.~N.~Aleixo, A.~B.~Balantekin and M.~A.~Candido Ribeiro,
{\it J.\ Phys.\ A} {\bf 34}, 1109 (2001)
[arXiv:quant-ph/0101024].

\bibitem{amado} R.D. Amado, F. Cannata, and J.-P. Dedonder, 
{\em Phys. Rev. A} {\bf 38}, 3797 (1988); {\em Int. J. Mod. Phys. A}
{\bf 5}, 3401 (1990). 

\bibitem{das} T.K. Das and B. Chakrabarti, {\em J. Phys. A:
Math. Gen.} {\bf 32}, 2387 (1999).

\bibitem{Balantekin:1997yh}
A.~B.~Balantekin and N.~Takigawa,
{\em Rev.\ Mod.\ Phys.}\  {\bf 70}, 77 (1998)
[arXiv:nucl-th/9708036].

\bibitem{jc} E.T. Jaynes and F.W. Cummings {\em Proc. IEEE} {\bf
51} 89 (1963).

\bibitem{shabat}
A.B. Shabat, {\em Inverse Prob.} {\bf 8}, 303 (1992).

\bibitem{Spiridonov:md}
V.~Spiridonov,
{\em Phys.\ Rev.\ Lett.}  {\bf 69}, 398 (1992)
[arXiv:hep-th/9112075].

\bibitem{Khare:gg}
A.~Khare and U.~P.~Sukhatme,
{\em J.\ Phys.\ A} {\bf 26}, L901 (1993) 
[arXiv:hep-th/9212147].

\bibitem{Barclay:1993kt}
D.~T.~Barclay, R.~Dutt, A.~Gangopadhyaya, A.~Khare, A.~Pagnamenta and 
U.~Sukhatme,
{\em Phys.\ Rev.\ A} {\bf 48}, 2786 (1993) 
[arXiv:hep-ph/9304313].

\bibitem{Fukui:1993xv}
T.~Fukui and N.~Aizawa, {\em Phys. Lett.} A {\bf 189}, 7 (1994).
[arXiv:hep-th/9309153].

\bibitem{Spiridonov:2003up}
V.~P.~Spiridonov,
arXiv:hep-th/0302046.

\bibitem{quesne}
C Quesne, {\em J. Phys. A} {\bf 35}, 9213 (2002). 

\bibitem{raman}
S. ~Ramanujan, {\em Messenger of Math.} {\bf 44}, 10 (1915); reprinted
in 
{\em Collected Papers of Srinivasa Ramanujan}, Ed. by G.H. Hardy,
P.V. Seshu Aiya, and B.M. Wilson (1927) (Cambridge University Press)
[reprinted by Chelsea, New York (1962)].

\bibitem{askey1}
R. ~Askey, {\em Amer. Math. Monthly} {\bf 87}, 346 (1980); {\em
  Applicable Anal.} {\bf 8}, 125 (1978/79). 


\end{thebibliography}
\end{document}